\documentclass[twoside]{dis08}
\usepackage[latin1]{inputenc}
\usepackage[dvips]{graphicx,epsfig,color}
\usepackage{wrapfig,rotating}
\usepackage{amssymb,amsmath,array}
\usepackage{cite}

\def\prp{\perp}
\def\prp{t}

\def\kt{\ensuremath{k_\prp}}

\def\as{\ensuremath{\alpha_s}}
\newcommand{\pom}{I\!\!P}
\newcommand{\xpom}{x_{\pom}}
\begin{document}
\title{What HERA may provide ?
 }

\author{Hannes Jung$^1$, Albert De Roeck$^2$, 
Jochen Bartels$^3$, 
Olaf Behnke$^1$, 
Johannes Bl\"umlein$^1$, \\
Stanley Brodsky$^{4,9}$, 
Amanda Cooper-Sarkar$^5$,  
Michal De{\'a}k$^1$, 
Robin Devenish$^5$, 
Markus Diehl$^1$,  \\ 
Thomas Gehrmann$^6$,  
Guenter Grindhammer$^7$, 
G\"osta Gustafson$^{2,8}$,  
Valery Khoze$^9$, 
Albert Knutsson$^1$, \\ 
Max Klein$^{10}$, 
Frank Krauss$^{9}$,  
Krzysztof Kutak$^1$, 
Eric Laenen$^{11}$,  
Leif L\"onnblad$^8$,   
Leszek Motyka$^3$, \\ 
Paul R. Newman$^{12}$, 
Fred Olness$^{13}$, 
Daniel Pitzl $^1$,
Marta Ruspa$^{14}$, 
Agustin Sabio Vera$^{2}$, 
Gavin P. Salam$^{15}$, 
\\
Thomas Sch\"orner-Sadenius$^{16}$, 
Mark Strikman$^{17}$ 
%
\vspace{.3cm}\\
%
1 - DESY, Notkestr 85, 22603 Hamburg, FRG\vspace{.1cm}\\
2  - CERN, CH-1211 Geneve 23, Switzerland\vspace{.1cm}\\
3 - University Hamburg, II Institut f\"ur Theoretische Physik, 22761 Hamburg, FRG\vspace{.1cm}\\
4 - SLAC, 2575 Sand Hill Road, Menlo Park, CA 94025-7090, USA\vspace{.1cm}\\
5 - Oxford University, Keble Road, OXFORD Ox1-3RH, UK \vspace{.1cm}\\
6 - Institut f\"ur Theoretische Physik, Universit\"at Z\"urich, CH-8057 Z\"urich, Switzerland \vspace{.1cm}\\
7 - MPI. Werner Heisenberg Inst. fur Phys, F\"ohringer Ring 6, D-80805 Munich, FRG\vspace{.1cm}\\
8 - Lund University, Department of Theoretical Physics, Solvegatan 14A, SE-223 62 Lund, Sweden \vspace{.1cm}\\
9 - Durham University, Inst. for Particle Physics Phenomenology, South Road
Durham DH1 3LE, UK\vspace{.1cm}\\
10 - University of Liverpool, Dept. of Physics, Liverpool L69 3BX, UK\vspace{.1cm}\\
11 - Nikhef theory group,    Kruislaan 409    1098 JS Amsterdam , The Netherlands \vspace{.1cm}\\
12 - University of Birmingham, Particle Physics Group, Edgbaston, Birmingham B15 2TT, UK\vspace{.1cm}\\
13 - Southern Methodist University, Dallas, TX 75275, USA \vspace{.1cm}\\
14 -  Universita' del Piemonte Orientale, via Solaroli 17, 28100 Novara, Italy \vspace{.1cm}\\
15 - LPTHE, UPMC Universit\'e  Paris 6 et 7, CNRS UMR 7589, Paris, France \vspace{.1cm}\\
16 - University Hamburg, Physics Department, Luruper Chaussee 149
22761 Hamburg, FRG\vspace{.1cm}\\
17 - Pennsylvania State University, Department of Physics,
University Park, PA 16802-6300
USA \vspace{.1cm}\\
}

\maketitle

\begin{abstract}
More than 100 people participated  in a
discussion session at the DIS08
workshop on the topic {\it What HERA may provide}. A
summary of the discussion with a structured outlook and list 
of desirable measurements and theory calculations  is given. 
\end{abstract}

\section{Introduction}

The HERA accelerator and the HERA experiments H1, HERMES and ZEUS stopped
running in the end of June 2007.
 This was after 15 years of
  very successful operation since the first collisions
in 1992. A total luminosity of $\sim 500 $ pb$^{-1}$ has been accumulated by
each of the collider experiments
H1 and ZEUS. During the  years  the increasingly better
understood and upgraded 
detectors  and  HERA accelerator have contributed significantly
to this 
success.
The physics program remains in full swing 
and plenty of new results were presented at DIS08
which are approaching the anticipated final precision,
fulfilling and exceeding the physics plans and the previsions of the upgrade program.

Most of the analyses presented at DIS08 were still based on the 
so called HERA I data sample, i.e. data taken until 2000, before the shutdown for 
the luminosity upgrade. This sample has an integrated 
luminosity of $\sim 100$ pb$^{-1}$, and the four times larger statistics
sample from HERA II  is still in the process of being analyzed.

Soon the LHC will start operation and will draw the attention of 
a large fraction of  HEP  physicists,
including many of those that have been working on HERA during the last years.
There is however still a lot to learn from the HERA data, much of it will be
of extremely high importance also for 
understanding LHC data, or for making  precision measurements 
 with LHC data. So it is worth continuing the analysis and 
investing the necessary time to complete this job.
Hence, it seems to be timely  to summarize and structure topics,
where analysis of HERA data  can contribute significantly
to answer
  general and fundamental questions 
of QCD. In order to benefit fully from the power of the HERA data
some further developments in theory are needed as well, which will be
addressed in this note too.

A discussion on future measurements at HERA 
took place at DIS 08, with more than 100
people attending a specially organized evening session. 
Many  interesting issues were brought
up. The clear support from the DESY directorate for fully utilizing
the HERA potential encouraged  us to prioritize a list of investigations and ideas
for what still can or should be done with HERA data. 
As reflected in this report, questions of factorization and universality
emerged as very prominent themes.

This paper is structured as follows: in the second section we discuss the
prospects of measurements of parton distribution functions (PDFs), which are 
the essential input for all calculations and predictions in hadron-initiated
processes. We also discuss the theoretical interpretation of such
measurements. 

When scrutinizing the emerging hadronic final state, issues like universality 
of hadronization, the color structure of the partonic final state and 
multi-scale phenomena become important. This is the topic of the third section.

Finally, since HERA is a machine perfectly suited for precision measurements 
in large areas of QCD, it is an ideal place to test the validity of the present 
theoretical approaches, and to limit their ranges. For example, deviations from 
linear evolution equations can be investigated, related to a regime where 
collective phenomena become dominant. Topics like this are discussed
in the fourth section.  

We discuss the impact of the measurements on tuning and 
validation of Monte Carlo event generators in the 
fifth section and give a conclusion in section 6.

In the final section we give a wish-list of measurements which should be 
done with HERA data.

Last but not least, one should secure the future. HERA has been the only 
lepton-proton collider in the world so far, and no construction of a 
similar machine is presently scheduled. Hence the data 
and analysis environment of the HERA experiments
should be preserved such that it is possible in future to turn to these data
for questions and studies, if  future  physics requires it.

\section{Parton distribution functions -- the irreducible input to all calculations
involving hadron beams}

The factorization theorems in QCD allow us to evaluate cross sections from
the convolution of corresponding hard scattering matrix elements, which are 
perturbatively calculable, with parton distribution functions (PDF), which 
include also non-perturbative physics. Although the factorization theorems 
are strictly proven only for a very limited number of processes, see for instance
~\cite{cteq-handbook}, their validity always is implicitly assumed when such
cross sections are calculated. Different factorization theorems exist, which 
are applicable to different regions of phase space:
\begin{itemize}
\item {\bf Collinear Factorization}, applicable at large enough $Q^2$.\\
	Cross sections are factorized into  process-dependent 
	coefficient functions $C^a$ and 
	collinear (integrated over $k_t$) parton distribution functions at a 
	factorization scale $\mu_f$ (e.g.\ $\mu_f^2=Q^2$ for the inclusive cross section
	in deep inelastic scattering):
	\begin{equation}
	\sigma = 
	\sigma_0 \int \frac{{\rm d}z}{z} C^a
	\left( \frac{x}{z},\mu_f\right) f_a\left( z,\mu_f\right)
	\label{collinear-factorization}
	\end{equation}
\item {\bf $\kt$ - Factorization},  applicable at small enough $x$ (high energy factorization).\\
	Cross sections are $\kt$--factorized~\cite{GLR,LRSS2,CCH,CE}
	into off mass-shell ($\kt$--dependent) partonic cross sections
	$\hat{\sigma}(\frac{x}{z},\kt)$ and  $\kt$--unintegrated parton 
	distribution functions ${\cal F}(z,\kt)$:
	\begin{equation}
 	\sigma  = \int\frac{dz}{z} \mbox{\rm d}^2\kt \;
	\hat{\sigma}\left( \frac{x}{z},\kt \right) {\cal F}\left( z,\kt \right)
	\label{kt-factorisation}
	\end{equation}
\end{itemize}
The proper and precise determination of the various PDFs is essential, as 
their uncertainties directly contribute to the uncertainties of the respective
cross section calculations.

Besides the determination of the parton distribution functions, the coupling
strength $\as(\mu^2)$ has to be determined. 
A 4-loop analysis of the non-singlet world data \cite{BBG}
resulted in
\begin{equation}
\alpha_s(M_Z^2) = 0.1141 \begin{array}{l} + 0.0020 \\ - 0.0022 \end{array}~.
\end{equation}
The complete statistics of HERA shall be analyzed in a corresponding
singlet analysis, from which a further improvement of the error of
$\alpha_s(M_Z^2)$ and improved error contours of the sea-quark and gluon distributions are expected. This analysis can be carried out at
3--loop order. Since the charm-quark contributions are large in the
small-$x$ region, the corresponding Wilson coefficients have to be calculated to the same level \cite{BBK1,BBK2,BBK3}.
Only if universal PDFs and $\as$ 
are measured and obtained can the whole factorization program be used in practice. Therefore 
the most precise measurements of these quantities relates to very fundamental
questions of our theoretical understanding.  Precision charged current
(CC) and  neutral current
(NC) cross-sections at high $Q^2$ also provide constraints on the electroweak couplings of the light quarks.

To extract the different PDFs, cross section measurements need to be performed
for various processes. The most precise of these measurements are obtained for 
the inclusive total cross sections. 
In semi-inclusive measurements (like jet, charm or 
diffractive cross sections), which 
require tagging of one or more particles in the final state, only parts of the phase space are experimentally 
accessible. Charmed - or diffractive structure functions, $F_2^c$ or $F_2^D$, 
can only be obtained after extrapolation from the visible to the total 
(partially invisible) phase space, which requires assumptions and thereby 
introduces an additional model dependence and systematic uncertainty.
For example, in the case of charm production, large model uncertainties enter 
through the extrapolation from the visible cross section of $D^*$ production in 
DIS to the total charm production cross section~\cite{F2charm-H1}. Measurements based on
the impact parameter have much smaller extrapolation uncertainties \cite{F2charm-impact-param}.

Therefore, aiming for minimal model dependence, emphasis has to be put on the 
measurement of mostly visible cross sections. In addition, in order to further
reduce the uncertainties in the measurements and thus of the extracted PDFs, 
efforts have already been started to combine the measurements of H1 and ZEUS.
This requires a detailed agreement concerning the kinematic 
ranges and accessible phase space. The combination of the inclusive cross section
 measurements~\cite{F2-H1+ZEUS}
of H1 and ZEUS, based on part of the HERA I data, is already leading to a 
very good precision. 
A combination of inclusive cross sections 
with measurements of heavy quarks and jets~\cite{Chekanov:2005nn} can be used to further constrain the 
gluon distribution function and $\as$. 
Also for the inclusive diffractive cross section a much better precision can be achieved from a 
combination of the  measurements from H1 and ZEUS, which has  already started.

Strategies to extract the most precise PDFs to be used at the LHC, and 
questions on how to use future LHC data to further constrain the 
PDFs are discussed in the PDF4LHC\cite{pdf4lhc} forum.

From these measurements the PDFs can be determined, either in the collinear and/or
the $\kt$-factorization approach. The combined measurements  constrain much better the 
PDFs, especially  the low $x$ gluon density \cite{H1+ZEUS-fits}. This new parameterization (HERAPDF) will be made available in LHAPDF.
An issue which is still open is whether the 
proton contains an intrinsic charm component. In order to verify and potentially quantify 
this experimentally, extra efforts to tag  
charm  in the very forward proton region would be needed.

However to fully appreciate the impact of highly precise measurements, further theoretical 
progress is required in the following issues:
\begin{itemize}
\item collinear factorization
\subitem $\circ$ inclusive cross sections and structure functions at three loop order in $\as$
\subitem $\circ$  heavy quark cross sections: transition from the massive to the
	massless approach
\subitem $\circ$ discontinuities when going beyond NLO in collinear factorization
\subitem $\circ$ higher order calculations for semi-inclusive cross sections (i.e. jet cross section at NNLO)
\item $\kt$-factorization
\subitem $\circ$ NLO calculations
\subitem $\circ$ unintegrated PDFs also for medium and large $x$ 
\item usage of PDFs in Monte Carlo event generators (PDF4MC)
\end{itemize}

Activities in these areas have started, but are not fully completed yet.

\section{Universality of the hadronic final state}

The hadronic final state in $ep$ collisions is much more complicated than in 
$e^+e^-$ annihilation, due to the presence of the colored proton remnant. 

Although it implicitly has been assumed for all theoretical predictions up to date,
the universality of parton-jet correlations at perturbative scales has never 
systematically been investigated in processes where more than one jet is produced. 
This correlation depends on the color structure of the final state, therefore
the so-called underlying event, which includes everything except the lowest-order 
process, might play a significant role. In addition, in the soft and 
non-perturbative regime the universality of hadronization and the parameters 
of the phenomenological hadronization model still waits to be verified in detail. 

\subsection{Color structure of final state}
The color flow between the partons of the hard process with the proton-remnant is responsible 
for the production of the hadrons in the phase space region between the hard scattering and the remnant. In diffractive 
events, the  color flow from the proton to the hard scattering is broken, 
in a simple model realized by the exchange of two gluons which neutralize each other's color. 
Single parton exchange (non-diffractive) and double parton exchange are only two extreme 
cases. Multi-parton exchanges, which are not in a color singlet state, are also possible 
and can increase the hadron multiplicity, since each of them will radiate further 
partons, possibly modified by interference effects. Multiparton interactions have been  studied with jets in photoproduction~\cite{Aid:1995ma,ZEUSmpi,h1-mpi-gp} and 
there are indications for interesting effects in DIS~\cite{h1-mpi-dis}.

In general, everything except for the lowest order process under investigation contributes to 
the so-called underlying event (UE). The UE includes contributions from initial and final 
state parton showers, as well as hadronization but also multi-parton exchanges including 
diffraction. Thus the study of the UE turns out to be as important and interesting for the 
basics of QCD as the hard perturbative processes itself are. 

Measurements of the transverse hadronic energy flow can be used to test the
\begin{itemize}
 \item universality of the color connection between hard partons 
 \item universality of the color connection between hard partons and proton-remnants
 \end{itemize}

The application of
hard scattering QCD collinear factorization \cite{collins} to 
the leading $Q^2$ component of diffractive DIS 
($ep \rightarrow eXp$) leads to
a concept of `diffractive
parton distribution functions' (DPDFs),
describing interactions which produce a leading final state proton with a
particular four-momentum.
Several authors have recently analyzed diffractive DIS data
and extracted DPDFs \cite{H1:F2D3,mrw,H1:difjet}. Many tests of the
factorization properties of diffractive DIS have been made by
comparing predictions based on these DPDFs
with observables from the diffractive hadronic final state, 
such as jet and heavy quark cross sections. 

Testing the factorization properties of diffraction and
improving on the precision of the DPDFs remains a major theme in diffraction
at HERA, with the diffractive longitudinal structure function $F_L^D$ a notable
new observable which may provide an interesting test of the large gluon 
density. However, perhaps the most important remaining work to be done
centers around understanding the manner in which diffractive factorization 
can be broken. There are two main themes, which are briefly discussed below.

As expected \cite{collins}, diffractive
factorization breaks down spectacularly when DPDFs from HERA
are applied
to diffractive $p \bar{p}$ interactions at the Tevatron \cite{cdf:dif}.
However, with the introduction of
a `rapidity gap survival probability'
factor to account for secondary interactions between the
beam remnants \cite{Dokshitzer:1991he,Bjorken:1992er}, a good description has been recovered. 
There remains much to be learned about gap destruction and survival, which
together with the HERA DPDFs are the two essential ingredients for predicting
the phenomenology of diffraction at the LHC. HERA may still
contribute here through the study of the onset of gap destruction events
in resolved photoproduction. 

A factorization - breaking effect which is present even in diffractive
DIS arises due to the presence of perturbatively calculable
configurations from $Q^2$-suppressed non-leading twist
$q \bar{q}$ fluctuations of longitudinally polarized
photons \cite{bekw}.
Although $q \bar{q}$ dipole scattering 
is not the dominant feature of inclusive diffraction at HERA, 
it gives rise to a good description of vector meson production, which
have a large component arising from longitudinally
polarized photons and are suppressed with increasing $Q^2$. 
Completing the program of study of exclusive processes will lead to a 
better understanding of the scattering of $q \bar{q}$ pairs from the proton. 
Non-factorizing longitudinal $q \bar{q}$ contributions also play a significant
role in the inclusive
cross section at large momentum 
fractions $z_{_{I\!\!P}}$ \cite{hebecker:teubner,mrw}, which must be better 
understood in order to decrease the large DPDF uncertainties in that region. 
This is 
essential for LHC preparations, as high $z_{_{I\!\!P}}$ inclusive processes
are the dominant background to 
central exclusive processes such as diffractive Higgs production. Among
other possibilities, 
a comprehensive search for exclusive contributions to diffractive dijet 
production is urgently needed.

\subsection{Universality of parton-jet relations}

The hard perturbative process in $e^+e^-,\;ep,\; pp $ collisions can be studied 
either by tagging the produced heavy objects (typically quarks in $ep$ collisions) 
or by measuring jets, which originate from hard partons produced in the hard
process. The jets are reconstructed from the observed hadrons. The precise 
definition of a jet, i.e. which hadrons and how they are combined into a jet depends on the jet 
algorithm and its resolution scale~\cite{Chekanov:2006yc}. It has been a long-standing issue that 
jet algorithms need to be infrared- and collinear-safe (i.e.\ the result must 
not change when a soft or collinear particle is added). Much progress has been 
made recently in providing also infrared- and collinear-safe definitions of 
cone-type jet algorithms \cite{siscone}. In all jet 
algorithms decisions are made on
 how many and which particles are clustered into a jet, and 
thus are potentially affected by the contribution from particles which do 
not 
belong to or originate from the hard perturbative process under consideration.

With a proper jet definition, the measured jet cross-sections can be compared with 
theoretical predictions calculated at fixed order at parton level or supplemented 
with parton showers (resummation) and hadronization. Similarly to the PDFs, the 
parton shower resummation is assumed to be universal. However, no explicit 
factorization theorem exists, which is applicable to the parton showers 
(although some steps in this direction have been made \cite{collins-stasto-rogers}), 
and thus experimental tests are vital.

Processes with one or two jets can provide experimental tests of
\begin{itemize}
\item universality of parton-jet correlation 
\item universality of soft gluon resummation 
\item universality and  importance of small $x$ resummation 
\end{itemize}

New and additional measurements are needed for a precise investigation of the 
parton-jet to hadron-jet relation for multi-jet events; especially the influence of multiparton interactions needs to be understood.

In the context of high energy factorization and the BFKL approach,  
forward jets well separated in rapidity from the outgoing lepton in 
DIS~\cite{Vera:2007dr,Kepka:2006xe} are sensitive to
next-to-leading order effects in the gluon Green's function, which have proved 
to be very important. In particular, the azimuthal angle correlations 
increase when higher order corrections are introduced (opposite to what is expected in 
a fixed order calculation) for a fixed value of 
$x$, while the jet and the electron become more
de-correlated as we increase the center-of-mass energy. 
Such  investigations  are important for the description 
of Mueller-Navelet jets at the LHC, and contribute to the production of 
high-multiplicity events~\cite{Bartels:2006hg}, where differences to
the standard approaches~\cite{Aurenche:2008dn} could be observed.

\subsection{Universality of hadronization}
Most of the parameters needed to describe the transition from on-shell or almost on-shell partons to observable 
hadrons have been precisely determined at $e^+e^-$ colliders, especially at LEP. However, 
the structure of the final state of events in $e^+e^- \to hadrons $ is different 
from that in $ep  \to  e' + hadrons $ and even more so from that in $ p p \to hadrons $. 
There are hints from HERA measurements that there are some differences of hadronization
in those process. For example, the $K/\pi$ ratio in $ep$-collisions seems to be different from 
that obtained in $e^+e^-$-annihilation events~\cite{k-pi-hera}. It is therefore crucial to
investigate the universality of hadronization --- the factorization of the soft and the 
perturbative region. 

A comparison of fragmentation parameters obtained at HERA with those from LEP will clarify 
the question of universality of hadronization. The measurements from HERA are important, 
as they allow us to understand the transition from  $e^+e^-$  to hadron colliders and they 
may be necessary for a precise  modeling of the hadronic final state at LHC.

Fragmentation functions are among the simplest quantities to describe the hadronization 
process in QCD.  
Measurements at HERA of {\it identified}  inclusive
hadron production would provide a strong impact to global fits. Furthermore, the $p_T$ spectrum and azimuthal distribution 
of the produced hadron provides insight into the intricacies of the dynamics and can be 
interpreted within a powerful theoretical formalism \cite{Bacchetta:2008xw}.

 $D$-meson production at HERA is fairly well described by NLO
calculations treating charm as massive and employing a fragmentation
function.  Measurements of distributions in $x_{Bj}$,
pseudorapidity, $p_T$ etc, have been performed by both H1 and ZEUS
collaborations. The dominant uncertainty in the NLO description is due
to the value of the charm (pole) mass, and a parameter in a charm
fragmentation function (e.g. Peterson, or Kartvelishvili). For a given
choice of PDF set and hadronization fraction $f(c\rightarrow D_i)$,
the value of these universal parameters
can be determined in a correlated fit. A first good attempt yielding
promising results can be found \cite{schagen}.

More measurements are needed for  a precise determination of fragmentation and 
hadronization parameters. With these measurements a detailed comparison of hadronization parameters and fragmentation 
functions with those obtained in $e^+e^-$ at LEP can be performed, and for the first time 
a systematic test of the universality of hadronization can be achieved.

\section{Deviations from expectations in linear QCD evolution equations}

Most of the topics discussed in the previous sections are related to single-parton 
exchange processes and to linear evolution equations, i.e.\ DGLAP, BFKL and CCFM. The
latter describe a regime which is normally classified as a dilute region, where the 
density of partons can be large, but still small enough such that they do not 
significantly interact with each other. However, when the density of partons becomes 
larger, they can shadow  or start to overlap and thus interact with each other. This is the regime 
of a dense system, where non-linear evolution equations are relevant, the oldest example 
of which is the GLR-MQ equation~\cite{GLR,Mueller:1985wy}. Hints at the onset of a dense region 
come from the observation of diffractive events at HERA and their interpretation 
in terms of the dipole picture. 
A strong signal for the onset of a dense region (or black disk limit) are  the measured diffractive
gluon PDFs, which indicate~\cite{Frankfurt:1999wk} that for $x\sim 10^{-4}$ and  $Q^2\sim 4 \, $
 GeV$^2$ the probability of diffraction in gluon induced processes reaches $\sim 40 \% $ which is close to the black disk limit of 50\%.
Now, the theoretical understanding of the dense region 
has received much support from measurements at RHIC, and  new evolution equations 
(like the  Balitsky-Kovchegov equation \cite{BK1,BK2} (BK)), which include non-linear terms,  
are available. However, the BK equation is derived for a large nucleus and only approximately
applicable to $ep$ and $pp$.  

Although saturation is theoretically well motivated, a clear, clean and indisputable experimental 
signature for it is still missing. To decide if and where nonlinear dynamical 
effects become important at HERA is difficult, especially since some signatures of 
saturation can be mimicked approximately within the linear DGLAP or BFKL descriptions.

However, the $\kt$-dependence 
of the unintegrated PDF as a function of $x$ could provide important information. 
In a linear scenario (BFKL) the parton
density (for fixed $\kt$) 
is expected to increase with deceasing $x$, while in the case of saturation this 
density will first increase, then flatten for smaller and smaller $x$ and will eventually decrease. As a function of  $\kt$ the parton density is expected to decrease for $\kt$ below the saturation scale and the $x$-dependence of the saturation scale can thus be studied directly. 
High-precision data in a wide kinematic region for dedicated observables 
will certainly help.

On the theoretical side progress is needed in
\begin{itemize}
\item the calculation of the evolution of unintegrated PDFs in the presence of saturation
\item the factorization and factorization breaking in the presence of saturation
\item the calculation of the change of the leading pion spectrum expected due to the onset of the saturation regime compared to the factorization prediction
\end{itemize}

Besides investigations on saturation, the range of validity of the linear evolution 
equations is not yet fully understood:
\begin{itemize}
\item in the moderate $Q^2$ region  contributions from higher twist effects (multi-parton exchange
	processes) are expected. However they are suppressed by additional powers of $1/Q^2$ 
	and therefore typically have only a small effect. At small $x$ this contribution 
	is increased by large $\log(1/x)$ terms. A systematic investigation of the 
	higher-twist region would require measurements at the same $Q^2$ but with $x$ varying
	over a larger range than available up to now. This can be achieved with $F_2$ measurements 
	 recorded at lower center-of-mass energies. 
\item in the large $x$ region a breakdown of the collinear factorization ansatz is expected 
	due to the transverse momentum as well as  energy momentum conservation as advocated in 
	~\cite{brodsky-largex}. 
\end{itemize}
The program to  investigate non-linear effects at HERA further and to constrain the validity 
of linear evolution equations is essential for any proper interpretation of small-$x$ effects 
at LHC. HERA is the only place  where these effects can systematically be studied in a 
clean and controllable environment, i.e.\ where precision measurements are possible. The 
results of such a program will have direct impact on measurements at RHIC but even more at LHC, 
where deviations from linear dynamics (saturation and multi-parton interactions) are expected even for high $p_t$ processes \cite{higgs-w-mpi}.

\section{Tuning and validation of MC event generators and models at HERA for future colliders}

With precision measurements, as described in the previous sections, different models and 
calculations can be systematically scrutinized. In many cases it will be possible to
find parameter settings which describe specific sets of measurements very well, while
failing for  different sets of observables, making different parameter settings unavoidable. 
Such a situation is not at all satisfactory, as it indicates deficits in our understanding of the 
underlying physics.

Since many of the calculations are also applied and used at different colliders (most prominently
the LHC), the investigation of the range of validity of various models is essential for their
success. For a detailed investigation all measurements from HERA, but also from other collider 
experiments, need to be available in a computer-readable form, which automatically includes 
all necessary cuts and reconstruction algorithms. Such frames exist in form of HZTOOL and 
its successor RIVET~\cite{rivet}. 

For the major Monte Carlo event generators, which are applicable for $ep$ as well as for $pp$,
such as ARIADNE~\cite{ARIADNE}, CASCADE~\cite{CASCADE}, HERWIG~\cite{HERWIG65} and 
PYTHIA~\cite{PYTHIA64}, a number of tunable parameters can be obtained from the 
measurements described here. These investigations should include:
\begin{itemize}
\item parton showers
\subitem $\circ$ tuning of hadronization parameters
\subitem $\circ$ tuning and validation of parton shower resummation; validity range of  collinear and  $\kt$ factorized parton showers
\subitem $\circ$ significance of angular ordered parton showers 
\subitem $\circ$ significance of LL or NLL parton showers 
\item parton distribution functions for MC event generators
\subitem $\circ$ LO or NLO PDFs 
\subitem $\circ$ determination of dedicated PDFs for MCs (PDF4MC) 
\subitem $\circ$ unintegrated PDFs
\end{itemize}

It could also happen that some  parameters are not uniquely  tunable to describe all the 
measurements. Such a situation indicates the incompleteness and inconsistency of the ansatz 
used and would be of general interest. HERA with its QCD precision measurements may well be the 
only place for a long time  where such a global validation of the different models can be done 
in an environment with a hadron beam and a controllable probe. In a global validation also measurements from $p\bar{p}$, $pp$  and $e^+e^-$ have to be included.

\section{Wish-list for measurements}
In the following we list the measurements which are needed in our view  to complete the program outlined above.
\subsection{Parton Distribution Functions}

A summary of cross section measurements relevant for the determination of the proton PDFs 
is given in Tabs.~\ref{tab:pdfmeausremnts} and~\ref{tab:updfmeausremnts}. 
The measurements described here are either already done or are to be completed with the full statistics of HERA II. 
The inclusive cross section measurements can be used to determine integrated as well as unintegrated PDFs.
Using the  full statistics will be important for measurements of the visible cross sections, especially for multi-jets and heavy flavor tags. 
The determination of unintegrated PDFs is complemented by measurements of semi-inclusive cross sections like charm or jet cross sections. Measurements of $\Delta \phi$ and $p_t$ correlations can further constrain the $\kt$--dependence of the unintegrated PDFs (Tab.~\ref{tab:updfmeausremnts}). A extended kinematic range to smaller $Q^2$ and smaller $p_t$ as well as more differential data as a function of $p_t$ and the dijet mass would be desirable, requiring new measurements.

\begin{table}[htb]
\centerline{\begin{tabular}{l|r}
  \hline
\multicolumn{2}{l}{\it neutral current (NC) (inclusive) }            \\ [4pt]\hline\\
$F^{\gamma/Z}_2(x,Q^2)$ ($e^{\pm} + p \to e^{\pm} + X$)    & $\as$, sea quarks,gluon \\ [4pt]\hline\\ 
$F^{\gamma/Z}_L(x,Q^2)$   ($e^{\pm} + p \to e^{\pm} + X$)                  & $\as$, gluon \\ [4pt]\hline\\
$F_2^{\gamma/Z}(x,Q^2)$ ($e^{\pm} + p \to e^{\pm} + X$)   at large $Q^2$             & $d/u$ at large $x$\\ [4pt]\hline\\
 $xF_3^{\gamma/Z}(x,Q^2)$   ($e^{\pm} + p \to e^{\pm} + X$)                 & $q-\bar{q}$\\ [4pt]\hline\\
$\sigma_{vis}^{K*}(x,Q^2)$   ($e^{\pm} + p \to e^{\pm} + K^*+ X$)    & strange-quark  \\ [4pt]\hline\\ 
$\sigma_{vis}^{D*}(x,Q^2)$   ($e^{\pm} + p \to e^{\pm} + D^*+ X$)    & charm-quark, $\as$, gluon  \\ [4pt]\hline\\ 
$\sigma_{vis}^{jets}(x,Q^2)$  ($e^{\pm} + p \to e^{\pm}+ \mbox{ n-jets}+ X$)   & $\as$, gluon  \\ [4pt]\hline
\multicolumn{2}{l}{\it charged current (CC)   }            \\ [4pt]\hline\\ 
$F_2(x,Q^2)$    ($e^- + p \to \bar{\nu} + X$)               & $u+c$,  $\bar{d}+\bar{s}$     \\ [4pt]\hline\\ 
$F_2(x,Q^2)$    ($e^+ + p \to \nu +  X$)                     & $d+s$,  $\bar{u}+\bar{c}$     \\ [4pt]\hline
\multicolumn{2}{l}{\it neutral current (NC) (diffractive) }   \\ [4pt]\hline\\ 
$F^{D(3)}_2(x,Q^2,\xpom)$ ($e^{\pm} + p \to e^{\pm} + X + p $)  & diffractive quarks, gluon \\ [4pt]\hline
\end{tabular}}
\caption{Summary of measurements relevant for the determination of the
	parton distribution functions. The right column indicates which quantities can be constrained by the measurements. }
\label{tab:pdfmeausremnts}
\end{table}

\begin{table}[htb]
\centerline{\begin{tabular}{l|r}
\hline
\multicolumn{2}{l}{\it charm in DIS  }            \\ [4pt]\hline\\ 
$\frac{d^2 \sigma (ep\to e' D^* + X)}{d x\; dQ^2}$, 
$\frac{d \sigma (ep \to e' D^* + jet +X)}{d p_t^{D^*,jet}}$ 
 & $ \int d \kt \; {\cal A}(x,\kt,\mu) $  \\ [8pt]\hline
\multicolumn{2}{l}{\it charm and/or jets in DIS  }              \\ [8pt]\hline\\ 
$\frac{d \sigma(ep \to e' D^* + jet +X)}{d \Delta\phi_{12}^{D^*,jet}}$, 
 $\frac{d^2 \sigma(ep \to e' D^* + jet +X)}{d \Delta\phi_{12}^{D^*,jet} d\Delta \eta_{12}^{D^*,jet} }$
 & $ \int dx \; {\cal A}(x,\kt,\mu) $  \\ [8pt]\hline\\ 
 $\frac{d^2 \sigma(ep \to e' D^*(jet) + jet +X)}{d \Delta\phi_{12}^{D^*,jet} dx}$ 
 & $ {\cal A}(x,\kt,\mu) $ \\ [8pt]\hline\\ 
$\frac{d^2 \sigma(ep \to e' D^*(jet) + jet +jet +X)}{d \Delta\phi_{12}^{D^*,jet} dx}$,
$\frac{d^2 \sigma(ep \to e' D^*(jet) + jet +jet +X)}{d \Delta\phi_{13}^{D^*,jet} dx}$ 
 & $ {\cal A}(x,\kt,\mu) $ \\ [8pt]\hline
\end{tabular}}
\caption{Summary of measurements relevant for the determination of the
	unintegrated PDFs ${\cal A}(x,\kt,\mu)$. The indices label the jets ($D^*$) ordered in $p_t$.
	The right column indicates which part of the unintegrated PDF can be constraint by the measurement. }
\label{tab:updfmeausremnts}
\end{table}

The photoproduction of muon pairs~\cite{Gluck:1992tq}  is largely induced by quark-antiquark 
annihilation and is therefore directly sensitive to the quark 
distributions in the real photon, which were up to now 
measured only in $e^+e^-$ collisions from the photon structure function 
$F_2^\gamma$. A measurement at HERA could probe a different flavor 
combination than the one appearing in $F_2^\gamma$, thereby yielding  
a measurement of the flavor structure of the quark content of 
the photon. Such a measurement would have important implications for 
predicting photon-proton cross sections at the LHC, and photon-photon 
cross sections at a future linear collider. A new measurement of $F_2^\gamma$ would complete the structure function measurement at HERA.

\subsection{Color structure of the final state}
The color structure of the hadronic final state can be investigated with measurements of transverse energy flow and charged particle spectra over the widest possible phase space:
\begin{itemize}
\item Universality of the color connection between partons from the hard process  (new measurements)
\subitem $\circ$ The energy flow between the jets depends on the color connection of the 
	hard partons. In diffraction, 3-jet events should show the same string-effect as in 
	$e^+e^-$-annihilation into three jets, whereas in non-diffractive events the 
	particle flow between the jets is expected to be different. A new measurement in the same 
	phase space region of diffractive and non-diffractive events needs to be performed.
\subitem $\circ$ The energy flow between the jets is also influenced by contributions 
	of multi-parton exchanges.  Thus a comparison of 2- and 3-jet measurements in 
	photoproduction with those at large $Q^2$ in the same phase space region will show 
	the importance of multi-parton interactions in DIS. In the photoproduction region, a comparison
	between direct photon enhanced and resolved photon enhanced regions for the same jet
	phase space is also sensitive to multi-parton exchanges.
\item Universality of the color connection between partons from the hard process and the proton-remnant (new measurements in extended phase space)
\subitem $\circ$ In non-diffractive events the transverse energy $E_T$ and the hadron 
	multiplicity in the forward region does not decrease as expected from collinear 
	factorization~\cite{eflow-h1}. \\
	Measurements of $E_T$ and hadron multiplicity as a function of $\eta$, $x$ and $Q^2$ 
	are not well described by theoretical predictions
	\cite{lund_smallx,eflow-h1,multiplicities}. 
\subitem $\circ$ In diffraction the formation of a rapidity gap is directly related 
	to the color structure of the exchange. Rescattering effects  
	between the remnants may destroy the rapidity gap. This is directly related to the 
	observation of factorization breaking in diffractive di-jet production in the 
	low-$Q^2$ region \cite{diff-fact-breaking}. A systematic study of factorization breaking 
	using diffractive di-jets 
	in photoproduction and DIS
	 as a function of the diffractive variables $x_{\pom}$, $\beta$ and  
	$Q^2$ are indispensable. The understanding of  rescattering effects is essential for 
	central exclusive Higgs boson production at the LHC.
\subitem $\circ$ New measurements (also in diffraction) of energy flow, hadron 
         multiplicity but also jets especially 
	close to the rapidity gap are needed. If the diffractive exchange 
	emerges from the proton due to processes which take place at 
	non-perturbative scales
	(and thus, if diffraction can be incorporated into the 
	starting condition of the usual PDFs), then the rate of jets with large transverse 
	momenta (forward jets) close to the rapidity gap should be small. However, if 
	the diffractive exchange contains a hard perturbative component, the cross section 
	of diffractive forward jets should be sizable. 
\subitem $\circ$ The measurement of  transverse energy $E_T$ and hadron multiplicity in the 
	forward region will be an important ingredient for any uniform description of 
	non-diffractive, diffractive and multi-parton interaction events as provided by
	the AGK cutting rules. Such a new measurement would require further investigations on tracking 
	in the forward region.
\end{itemize}

\subsection{Universality of parton-jet relations}
The universality of the correlation between parton and jets (at parton or hadron level) can be investigated with processes which have one or two jets. The following, partially new, analyzes would be useful:
\begin{itemize}
\item correlation  between hard partons and jets at parton level 
\subitem $\circ$ The correlation can be investigated by measurements of the di-jet 
	production cross section as a function of the di-jet mass $m_{ij}$, thereby avoiding 
	to compare directly the transverse energy and rapidity of the jet with those of the parton jet in the calculation. 
\subitem $\circ$ The  correlation as a function of the jet $p_t$ as well 
	as the influence of the underlying event can be quantified by jet production
	cross section measurements from lowest possible $p_t \sim 1$~GeV to high $p_t$.
	The influence of soft and collinear radiation can be studied with measurements 
	as a function of the jet resolution parameter and as a function of the 
	jet algorithm (inclusive $k_t$,  anti-$k_t$, SiSCone, etc.). In addition,
	energy flows inside the jets or the jet profiles could shed light on this 
	effect.
\subitem $\circ$ Measurements  of 
	$\Delta p_t  =  |p_{t;1} - p_{t;2}|$ or $\sum p_t =  p_{t;1} + p_{t;2}$ for di-jets, with 
	$p_{t;1,2}$ being the transverse momentum of the jets, as a function of 
	$p_t^{jet}$, $\eta^{jet}$, $Q^2$ and $x$ are extremely important as they give 
 	direct access to higher-order contributions.
\item universality of soft gluon resummation \\
The probability to have two jets exactly back-to-back in the 
	$\gamma^* p$ center-of-mass should vanish since gluon radiation from the initial and final state 
	destroys the LO back-to-back configuration. This so-called Sudakov effect also 
	plays an important role for the $p_t$-spectrum of $W/Z$ production as well 
	as the $p_t$ distribution for the Higgs boson in hadron-hadron collisions. \\
	The effect of soft gluon radiation can be investigated by di-jet measurements 
	as a function of $\Delta \phi$, as a function of the di-jet mass and of
	$Q^2$, ranging from photoproduction to high $Q^2$. In the region of 
	$ \Delta \phi \sim 180^o$ contributions from soft gluon resummation can 
	be studied.
\item importance of small $x$ resummation \\
	At large enough energies or small enough $x$ the suppression of higher-order 
	contributions (due to higher orders in $\as$) is compensated by logarithms of
	$1/x$, thus the jet (or leading particle) cross section increases with 
	$p^2_t \sim Q^2$. At small values of $x$ also the transverse energy and the
	particle multiplicity become larger than expected from pure hadronization.  
	Small $x$ resummation can play an important role for the 
	$p_t$-spectrum of $W/Z$ production as well as the $p_t$ distribution for the 
	Higgs as pointed out in \cite{cteq-pt-w}.
	The small $x$ resummation can be studied by the forward jet cross section 
	associated with jets or heavy quarks in the central region. 
\subitem $\circ$ Important information on parton radiation can be obtained by a 
	measurement of the transverse energy flow as a function of $\eta$ in events 
	with a forward jet.
\subitem $\circ$ The measurement of the DIS di-jet cross section in the region of 
	$ \Delta \phi < 120^o$ is sensitive to contributions beyond $2 \to 3$-processes 
	and thus can signal effects of all order resummation. The transverse energy flow
	and particle multiplicity in DIS di-jet events as a function of $ \Delta \phi $ and jet $p_t$ is 
	important.
\subitem $\circ$ The measurement of forward jet production with $p_t^2 \sim Q^2$	can provide
          essential information on small $x$ resummation. The forward jet needs to be at smallest 
          possible angle w.r.t. proton beam axis. The cross section as a function of the angle $\phi$ 
          between the scattered electron and the forward jet, as a function of the rapidity separation 
          between both, and as a function of the jet $p_t$  is essential.
\end{itemize}
The relation between partons (or jets at parton level) and jets at hadron level can be investigated with the following new measurements: 
\begin{itemize}
\item the inclusive single jet production cross section as a function of the jet resolution 
	parameter $R$, as well as the differential jet cross section as a function of $x$, $Q^2$, 
	transverse momentum $p_t^{jet}$ and $\eta^{jet}$.
\item the investigation of the above cross sections for different jet algorithms 
	and a comparison with theoretical predictions  at parton level and with those after  
	parton showering and hadronization.
\item measurement of multi-jet production cross sections as a function of $N_{jet}$ in 
	photoproduction and DIS as a probe for the relevance of higher-order contributions.
	This measurement could be repeated using the full HERA II statistics.
\item measurement of differential multi-jet cross sections in photoproduction and DIS 
	for an investigation of higher-order contributions. Correlations between the jets 
	in $\phi$ and $p_t$ should allow one to separate multi-parton interactions from multi-jet 
	production coming from a single interaction (similar to what was done in \cite{gamma-jets}).
\item measurement of particle multiplicity and energy flow in multi-jet events. Underlying 
	event contributions and multi-chain processes will show an increasing activity 
	away in $\eta$ and $\phi$ from the hard process.
\item measurements of jet cross sections with equal and with very different transverse momenta 
	for an investigation of multi-scale processes.
\end{itemize}

\subsection{Universality of hadronization}
The effect of hadronization in general is difficult to separate from soft parton radiation. Energy flow, charged and neutral particle multiplicities can be used to determine general properties of hadronization, requiring new investigations:  
\begin{itemize}
\item   particle multiplicity and transverse energy distributions, differential in $x$, $Q^2$ as 
	well as in $\eta$ and $\phi$. This would also potentially provide insights into 
	the role of small $x$ dynamics (BFKL), multiparton radiation (higher order
	contributions and parton showers) as well as multiple-parton interactions.
\item particle multiplicity, energy and $p_t$ spectra in the forward region: this 
	would help to fix the fragmentation of the proton-remnant. Natural questions
	to ask include: Does the proton remnant fragmentation depend on $x$ or $Q^2$? 
	What is the intrinsic $k_t$ of quarks and gluons in the proton? HERA data could
	help to answer such questions.
\end{itemize}
The following measurements (using full statistics and extended phase space)
are needed for  a precise determination of fragmentation and 
hadronization parameters:
\begin{itemize}
\item measurements of charm fragmentation functions and at the same time the charm (pole) mass  as a function of $x$, $Q^2$, $p_t$ 
	and $\hat{s}$
\item measurement of non-strange meson ($\pi$) and baryon ($p$, $n$) and 
	strange meson and baryon production (hyperons) as a 
	function of $x$, $Q^2$, $\eta$, $\phi$ and $p_t$ and correlations between them, 
	needed in particular for the determination of light and strange hadron fragmentation functions 
\item quark and gluon jet fragmentation: measurement of leading and subleading 
	particles in jets to allow a separation of quark and gluon jets.
\item measurement of proton and anti-proton production in the forward region 
	(proton-fragmentation region)
\end{itemize}

\subsection{Deviations from expectations in linear QCD}

Deviations from linear QCD might be seen in inclusive measurements but the best observables are semi-inclusive measurements.  The following measurements using the full statistics could show effects of non-linear QCD:
\begin{itemize}
\item $\kt$-dependence 
         of the unintegrated PDF  for different values of $x$ with the transverse momentum of a di-jet pair, a 
         heavy-quark pair or  a $J/\psi +g $ system.
\item the energy dependence of the ratio $F_2^D /F_2$ of diffractive and inclusive 
	structure functions.
\item geometric scaling of the inclusive structure function$F_2 - F_2^{c}$
         (with the charm contribution subtracted)
\end{itemize}

\subsection{Proposals for additional measurements}
Here we present proposal for measurements in addition to the program outlined above. 
\subsubsection{Intrinsic Heavy Quarks at HERA \cite{brodsky-proposal}}

As emphasized by the CTEQ group~\cite{Pumplin:2007wg}, there are indications that the structure functions used to model charm
and bottom quarks in the proton at large $x_{bj}$ have been strongly underestimated, since they ignore the intrinsic heavy quark fluctuations of
hadron wave-functions. 
The probability for Fock states of a light hadron such as the proton to have an extra heavy quark pair decreases as $1/m^2_Q$ in non-Abelian
gauge theory~\cite{Franz:2000ee,Brodsky:1984nx}.  The intrinsic Fock state probability is maximized at minimal off-shellness; {\em i.e.}, when the constituents have minimal invariant mass and equal rapidity. Thus the heaviest constituents have the highest
momentum fractions and the highest $x_i$. Intrinsic charm thus predicts that the charm structure function has support at large $x_{bj}$ in
excess of DGLAP extrapolations~\cite{Brodsky:1980pb}; this is in agreement with the EMC measurements~\cite{Harris:1995jx}.

The SELEX~\cite{Russ:2006me} discovery of $ccd$ and $ccu$ double-charm baryons at large $x_F$ reinforces other signals for
the presence of heavy quarks at large momentum fractions in hadronic wave-functions, which is a novel  feature of intrinsic heavy quark Fock states.
This has strong consequences for the production of heavy hadrons, heavy quarkonia, and even the Higgs at the LHC. Intrinsic charm and bottom
leads to substantial rates for heavy hadron production at high $x_F$~\cite{Brodsky:2006wb}, as well as anomalous nuclear effects.  The heavy quark distributions in the proton at large $x$ are perhaps the most important uncertainties in hadron structure; the uncertainties in $c(x,Q^2)$ also causes confusion when one uses charm production to tag gluon distributions.

Although HERA measurements of charm and bottom cross sections in deep inelastic $e p$ scattering are normally restricted by kinematics and rate to small $x_{bj} $, there is some chance of seeing excess $\gamma^* p \to c X$ events at high $Q^2$ and $x$.   

In addition, other hard scattering reactions may allow access the intrinsic component of heavy quark distributions at large $x$ at HERA:
\begin{itemize}
\item Study the hard photoproduction process $\gamma p \to c X$ where the charm jet is produced at large $p_T$.  The dominant subprocess is $\gamma c \to c g,$ where a gluon jet recoils against the charm quark trigger.
\item Look for  open or hidden charm production at high $x_F$ in the proton fragmentation region in normal high $Q^2$ deep inelastic events, possibly using the existing forward silicon detectors.    In this case one looks at fast charm produced from the excitation of the $|uud c \bar c>$ Fock state of the proton.
\item Use $ e p \to e^\prime \gamma c X$  to effectively lower the electron energy; this would require tagging a forward photon along the electron direction.
\end{itemize}

\subsubsection{Exclusive Processes~\cite{khoze-proposal}}
\begin{itemize}
\item The extraction of the bare $3\pom$ vertex \\
The triple-Pomeron vertex is an important  ingredient
for the physics of diffraction~\cite{kmr1,kmr1b}. In order to determine
the value of the bare (unscreened) vertex, the triple-Pomeron
interaction should be measured in a process where the
rescattering  effects are  suppressed.
Such a process is proton dissociation in the inelastic diffractive
$J/\psi$ photo (or electro-) production. 
Unfortunately, the existing HERA data are  fragmentary and
no results on the distribution over the mass $M$ of the
system Y accompanying $J/\psi$ are available.
 We need better statistics of inelastic diffractive $J/\psi$
events. The inelastic diffractive $\Upsilon$ events are of a
special interest, but are limited by the recorded statistics.
It is crucially important to have the data with  an explicit
measurement of the proton dissociation mass spectrum-spectrum 
in order to perform the
full triple-Regge analysis which will allow to separate
the $\pom\pom\pom$ term from the other tripple-Regge contributions.

\item More precise measurements of exclusive $\Upsilon$
photoproduction \\
Exclusive $\Upsilon$ photoproduction allows to probe the generalized
unintegrated gluon distribution in a kinematical region which
is close to the expected one  in central exclusive production of the Higgs boson
at the LHC \cite {kmr2}.
Currently, the uncertainties caused by the lack of
knowledge of the gluon distribution at low $x$ and small scales
are sizable, and better statistics of exclusive $\Upsilon$ event will
allow to constrain the expectations for the central exclusive production
of new physics events at the LHC. Also exclusive $\gamma p$ collisions
at the LHC could  be directly used.

\item Measurement of the ratio  $R$ of diffractive and inclusive di-jet
photoproduction. \\
It was suggested in \cite{kkmr} that a good way to study the
effects of factorization breaking in diffractive di-jet photoproduction
is to measure the ratio $R$ of the diffractive process to the corresponding
inclusive production process. In this ratio many theoretical
and some experimental uncertainties  can cancel.
It will be very interesting to have the results  on $R$ as a function of $Q^2$
in order to observe variation of absorptive effects.

Further studies of the $E_T$ dependence of the screening effects in diffractive
di-jet photoproduction are very important.  It is expected (for instance \cite {kkmr1})
that with decreasing of the jet  $E_T$ the screening effects become stronger.
This is because at lower $E_T$ the role  of the large size
diffractive component with larger absorptive cross section  becomes more pronounced.
The new H1 and ZEUS data seem to indicate such behavior but more data are urgently needed.
\end{itemize}
\subsubsection{Semi exclusive diffraction~\cite{strikman-proposal}}
\begin{itemize}
\item Hard Pomeron trajectory \\
One of the important  issues of the hard Pomeron dynamics is the $t$-dependence of the hard Pomeron trajectory. It has been measured at HERA for small $t$ in the exclusive electro (photo)   production of onium states.
To study it at large $t$ it is necessary to use rapidity gap events at large $t$: $\gamma(\gamma^*) + p \to VM + \, gap \, X$. A number of such analyses were performed at HERA. However, practically in all cases $M_X^2/W^2$ was kept constant. It turns out that in this case sensitivity to $\alpha_{\pom}(t)$ is very low as the energy dependence is mostly given by $x$ dependence of the gluon density in the target \cite{Frankfurt:2008er}. It is necessary to perform analyses of the energy dependence of the process for the fixed upper limit on $M_X$. In this case sensitivity to $\alpha_{\pom}(t)$ will be maximal.
\item Improving knowledge of the transverse distribution of small $x$ partons \\
Knowledge of   transverse distribution of gluons and quarks (which is a Fourier transform of the 
$ t$-dependence of gluon and quark GPDs  measured in exclusive DVCS and DIS vector meson production) is crucial for  a realistic modeling of the geometry of the $pp$ collisions at the LHC. One needs to determine more precisely the difference between the $t$-slopes of DVCS, electro and photo production of $J/\psi$ and measure the slope of the $\Upsilon $ photoproduction.
\item Gluon fluctuations in the nucleon \\
It was demonstrated in \cite{FSTW} that the ratio $$R(W,Q^2,M_X)=\frac{\frac{d\sigma (\gamma^* +p\to VM+ M_X)(t=0)}{dt}}  
{\frac{d\sigma (\gamma^* +p\to VM +p)(t=0)}{dt}}$$ measures the variance of the fluctuations of the gluon density  in the nucleon for a given $x$. A slow $Q^2, x$ dependence of this ratio is predicted. 
Hence a systematic experimental study of this ratio will provide an important new information about structure of the
nucleon and will allow a better modeling of the $pp$ collisions at LHC.
\item Novel two $\to $ three processes\\
There exists a number of novel DIS processes which were not studied so far which are sensitive to the generalized PDFs (GPDs) in the nucleons and large $t$ GPDs in the nucleon. One example \cite{Strikman:2003gz}
is the process which maybe feasible for detection at HERA is 
the process $\gamma^*+ p\to VM + gap + \pi^+ + n$ where a pion is produced with $x_F\le 0.1, p_t\ge 1.5$~GeV/c  corresponding to relatively small rapidities and the neutron hits the neutron calorimeter  and has $x_F\ge 0.9, p_t\le 0.1$ GeV/c. The transverse  momenta of the VM and pions are nearly balanced in this kinematics and selection of large t for the process (color transparency)  and low $p_t$ for the neutron  lead to  a suppression of the final state interaction. This process is expected to be reasonably enhanced for large $t$ since it has a much weaker $t$ dependence than the  $\gamma^* +N \to  VM + N$ process (a factor $\propto 1/t^2$) due to a weaker large $t$ dependence pion GPD as compared to the nucleon GPD.
\end{itemize}
\subsubsection{Observation of the Odderon at HERA}

Odderon exchange has never been observed in experiment,  even though it is an essential prediction of QCD.

The asymmetry in either the fractional energy distribution or the  angular distribution of charm versus anti-charm jets produced in high energy diffractive photoproduction 
$$\gamma^* p \to c \bar c p$$ 
is sensitive to the interference of the Odderon $(C = -)$ and Pomeron $(C = +)$ exchange amplitudes in QCD~\cite{Brodsky:1999mz}. 
This asymmetry has been estimated to be of the order 15\% using an Odderon coupling to the proton which saturates constraints from proton-proton vs. proton-antiproton elastic scattering. 

Measurements of this asymmetry at HERA could provide firm experimental evidence for the presence of Odderon exchange.

\section{Conclusion}

HERA was and is still the only place where precision measurements in QCD in a controlled 
environment with an electron probe and  a hadron beam are performed. The vast amount of data collected at the  
HERA II collider with excellently  understood 
detectors is still in the process of being analyzed.

Many fundamental questions of QCD, such as the 
  universality of the PDFs for different processes, 
hadronic final state and hadronization universality
can be addressed with HERA data. The answers to these questions 
are extremely interesting  on their own, but even more so, this  
will be  necessary 
ingredients for many potential discoveries at the LHC and elsewhere.

It has been discussed in this note that  the HERA physics program is still
extremely rich, full of potentially very important measurement
in many different areas of QCD.  We tried to structure 
the topics and to focus on the most fundamental questions 
such as universality 
and factorization. 

We believe that much progress can be achieved in theses fields in the next few
years, and we look 
forward to many exciting and challenging new results.

\section{Acknowledgments}
We would like to thank the whole organizing committee of DIS08 for this very inspiring and interesting conference. We are especially grateful for the possibility to hold this discussion session and for the invitation to contribute with this note to the DIS proceedings.

A very special thank-you goes also to all the participants of this evening discussion session, which was so lively and productive  that we were encouraged to write up this note. This session showed that HERA physics is interesting also for young physicists and that QCD is still a challenging field, with many ups and downs but clearly the potential for interesting discoveries.
 

\begin{footnotesize}
\raggedright

\end{footnotesize}


\end{document}